# COMBINING THE FOGG BEHAVIOURAL MODEL AND HOOK MODEL TO DESIGN FEATURES IN A PERSUASIVE APP TO IMPROVE STUDY HABITS

**Justin Filippou**
School of Business IT & Logistics
RMIT University
Melbourne, Australia
Email: j.filippou@student.rmit.edu.au

**Christopher Cheong**
School of Business IT & Logistics
RMIT University
Melbourne, Australia
Email: christopher.cheong@rmit.edu.au

**France Cheong**
School of Business IT & Logistics
RMIT University
Melbourne, Australia
Email: france.cheong@rmit.edu.au

## Abstract

Using technology to persuade people to behave in a certain way is an ever-increasing field of study. The ability to persuade individuals is quite clear in e-commerce, where individuals are persuaded to make purchasing decisions. However, it can also be applied to other disciplines, such as education where improving the study behaviour of students would be particularly useful. Forming good study habits can be a challenge for university students who have not done so in the earlier years of their education, or where the pressures of external commitments have eroded previously good habits. We use a combination of the Fogg Behavioural Model and the Hook model to design features for an app as a component of a larger persuasive system to help improve three key areas of study habits: study scheduling, class preparation and group study. The app will be built and tested in a university setting targeting undergraduate students.

## 1   Introduction

As computing technology becomes more powerful and mobile, it is becoming further engrained into peoples' daily lives and influencing how they behave. In fact, technology can now be designed for the exact purpose of influencing behaviour and has been termed "persuasive technology" (Fogg 2002; Hamari et al. 2014; Kaptein et al. 2009; Lockton et al. 2010). For instance, understanding and influencing the buying behaviour of customers is highly sought after by e-commerce companies (Saari et al. 2004). By understanding the typical purchasing patterns of their customers, businesses can tailor specific suggestions for future purchases and increase the likelihood of repeat business (Chen and Popovich 2003). Take for example Amazon[1], which displays other related items when a customer is about to pay for their items to purchase. The customer may know of the recommended product but would otherwise not have considered purchasing it. However, by suggesting it at the time of purchasing a similar item, the user may be influenced to buy. An example of how mobile technology can be used effectively is the eBay[2] app. Users can bid on items through the app and are given push notification reminders when auctions are about to end or have been outbid, prompting them to re-focus their attention on eBay. Both Amazon and eBay have been very successful in forming customer habits of using their services regularly, in part due to their implementation of persuasive technology.

---

[1] www.amazon.com
[2] www.ebay.com

The use of persuasive technology is not simply limited to the e-commerce. Another area where persuasive technology can be appropriate is in education, particularly for tertiary students. Influencing the study behaviour of tertiary students has traditionally been very difficult. This is because they have formed study habits (both positive and negative) from a young age in the early years of school. Much like a consumer being aware of a product but not purchasing it, students often are aware of study habits that would lead to improvements in learning outcomes, but still choose not to perform the more beneficial behaviour. Therefore, technology intervention may be required in order to overcome these difficulties. In this research, an existing technology-based teaching solution is used as a platform in which new persuasive features will be incorporated.

## 2　Background

This research is part of a broader project to develop a system to persuade students to improve their study habits. To achieve this, the system needs to be developed in accordance with established theories regarding persuasive technology. This section discusses the theoretical aspects to consider when designing such a system. Firstly, study habits will be discussed as the focus of this research is to identify important study habits and then encourage them to be adopted by users by designing persuasive features.

### 2.1　Study habits and academic performance

Study habits form an integral part of overall academic performance (Credé and Kuncel 2008). For many students, establishing and maintaining positive study can be difficult. This can be due to both knowing what good study habits are, but being unable or unmotivated to do so, and not being aware of what positive study habits are. To alleviate this, a persuasive system can be designed to support students and assist them with forming the positive study habits. However, there are many study habits that a student could have (Pintrich 1991). Given the ultimate goal of improving study habits is to improve academic achievement, we need to identify the most important study habits for academic performance. This was carried out in previous work, which led to the development of three models to measure the impact study habits have on different dimensions of academic achievement: self-perceived performance, objective performance (receiving grades) and performance change over time (Filippou et al. 2015). The study surveyed current students and graduates using an online questionnaire which was adapted from the Motivated Strategies for Learning Questionnaire (MSLQ) (Pintrich 1991; Pintrich et al. 1993). The resulting models provide a list of potential habits (refer Table 1), which we can target when building persuasive educational technology.

| Academic Dimension | Model |
|---|---|
| *Self-perceived performance* | ***f(x) = How would you describe your academic performance as a student?*** <br><br> $f(x) = (0.18)x_1 + (-0.21)x_2 + (-0.28)x_3 + 4.39$ <br><br> $x_1$ = *When I study for a class I pull together information from different sources, such as lectures, readings and course materials\** <br><br> $x_2$ = *I often get so lazy or bored when I study for a class that I quit before I finish what I planned to do\*\** <br><br> $x_3$ = *When a subject's work is difficult, I either give up or only study the easy parts\*\** |
| Objective performance | $f(x) = $ **How often did you receive high grades (over 80%) for assignments, exams or subjects overall?** |
| | $f(x) = (0.24)x_1 + (0.30)x_2 + (-0.19)x_3 + (-0.19)x_4 + (0.14)x_5 + 2.16$ |
| | $x_1$ = When I study for a class, I pull together information from different sources, such as lectures, readings, and discussions** |
| | $x_2$ = I usually study in a place where I can concentrate on my work** |
| | $x_3$ = I find it hard to stick to a study schedule** |
| | $x_4$ = It is my own fault if I don't learn the material in a subject* |
| | $x_5$ = When I study for a subject I write brief summaries of the main ideas from the readings and my class notes* |
| Performance change over time | $f(x) = $ **How did your learning performance as a university student change over time?** |
| | $f(x) = (0.23)x_1 + (0.27)x_2 + (0.18)x_3 + (-0.30)x_4 + (-0.25)x_5 + (0.30)x_6 + 2.75$ |
| | $x_1$ = I try to change the way I study in order to fit the subject's requirements and the instructors teaching style** |
| | $x_2$ = When I study for a subject, I often set aside time to discuss material with a group of students from the class** |
| | $x_3$ = When I study for a class, I practice saying the material to myself over and over* |
| | $x_4$ = During class time I often miss important points because I'm thinking of other things*** |
| | $x_5$ = When studying for a subject, I often try to explain the material to a classmate or friend** |
| | $x_6$ = I'm confident I can understand the most complex material presented by the instructor in a subject** |

*Note: * p < .05, ** p < .01, *** p < .001*

*Table 1. Summary of important study habits for academic performance (reproduced from Filippou et al. (2015))*

## 2.2 Phases of Designing Persuasive Systems

Persuasive system design can be broken down into three distinct phases: (1) understanding key issues behind the persuasive system, (2) analysing the persuasion context and (3) designing the system qualities (Oinas-Kukkonen and Harjumaa 2009). The first phase is concerned with understanding that technology is no longer neutral and that inherently, there is some level of persuasion. However, this research addresses the second and third phases as the first phase has already been carried out by identifying the most important study habits as described in section 2.1.

### 2.2.1 Persuasion Context

The second phase of designing persuasive systems is concerned with understanding the persuasion context. The persuasion context consists of three aspects: the *intent*, the *event* and the *strategy* (Oinas-Kukkonen and Harjumaa 2009). The *Intent* involves understanding who the *persuader* is and what the desired *change type* is. The change type refers to whether it is attitude or behaviour change that is the ultimate goal. Secondly, understanding the event involves analysing what happens around the behaviour or habit to influence. This must also be carried out with the user's goals in mind. If the goals of the system and the goals of the user do not align, the likelihood of persuasion occurring is hampered. Finally, the strategy for persuasion needs to be made clear, which is about understanding the *message* that will be sent that will encourage the persuasion, and how it will be done. These three key areas form the high-level theoretical aspect to designing persuasive systems. Once these areas have been addressed, designing the system can proceed to the actual system features. This is the third phase of persuasive systems development.

### 2.2.2 Designing System Qualities

The third phase of persuasive system development is the design of the system qualities. For technology to be useful to people it needs to support them in achieving their desired goals. It is possible to classify support in four different ways. Those being: *primary task support*, *dialogue support*, *system credibility support* and finally *social support* (Oinas-Kukkonen and Harjumaa 2009). *Primary task support* is where the system helps the user carry out the main task they are undertaking by reducing the effort required to perform the task, guiding the user sequentially through steps, tailoring information about the task, personalising content, tracking progress, simulating cause and effect or through rehearsal of the target behaviour. *Dialogue support* involves providing the user with direct feedback such as praise, rewards, reminders or suggestions to perform the target behaviour. *System credibility support* is important to establish trust with the user. To do this, the system should demonstrate expertise, experience and competence in the subject area and have a competent look and feel. Finally, a system can improve persuasiveness by incorporating *social support*. Humans are naturally social beings, and so providing an opportunity to work with the other users to help alter their behaviour and compete with them to motivate the change to occur. This is particularly useful if users with similar goals are brought together.

## 2.3 Designing System Features

Although phase three of the design process outlines the theoretical underpinnings of the system qualities, it doesn't provide guidance about how the actual operation of the features should work. To assist with this, the Hook model for building habit-forming products may be used (Eyal and Hoover 2013). The name of model derives from its ability to "hook" people in to using a (typically computer-based) product and ultimately form a habit. There are four components that enable this to occur. Firstly, a user needs to be *triggered* to perform an action either from an internal or external cue. Initially, the user is more likely to require an external trigger, as there is minimal internal desire to perform an action. Secondly, the user will need to perform an *action*, which is a small amount of work in anticipation of reward. Next, the user needs to be *rewarded* for performing the previous action. The reward is not required immediately and can be done in a variable amount of time. The variability of random reward also provides stronger positive reactions to the reward. Finally, the system needs to allow the user an opportunity to have some sort of investment in the system, which will increase the likelihood of the user returning. By investing some time or effort into the system, the user has established a connection to the system and will want to return. This process then happens again in a cyclical manner, as illustrated in Figure 1.

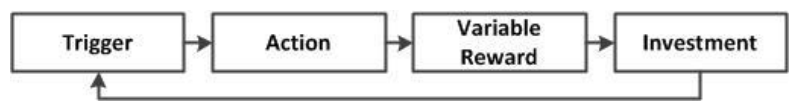

*Figure 1: The Hook model (adapted from Eyal and Hoover (2013))*

An example of how the Hook model has been used in commercial web applications is apparent in how LinkedIn[3] functions. When a user initially joins LinkedIn, the web site displays an empty progress bar indicating that their profile needs to be completed. This is an external trigger. The user then clicks on the link to improve their profile, which is an example of an action being performed. After some time has passed, the user is sent an email highlighting that their profile has since been viewed by many people because of the details they provided, rewarding the user for their action. The user may then elect to make a connection with the other users, indicating that an investment has been made by the user and therefore is likely to return.

A limitation of the Hook model is that it overlooks some critical aspects that contribute to the likelihood of triggers being effective, with those being the ability and motivation of the individual to perform the desired action. One model that explains the relationship between *motivation, ability* and *triggers* is the Fogg Behavioural Model (FBM) (Fogg 2009a). The FBM suggests that if performing a behaviour is difficult to do and there is little motivation to do so by the individual, then an external trigger will not be effective. If ability is low but motivation is high, then the trigger will only be effective if the system can support the user to reduce the difficulty in some manner. If motivation is low and ability is high, then the system needs to provide some form of incentive to improve motivation first, before a trigger will work. This relationship can be easily represented graphically (see Figure 2).

---

[3] http://www.linkedin.com

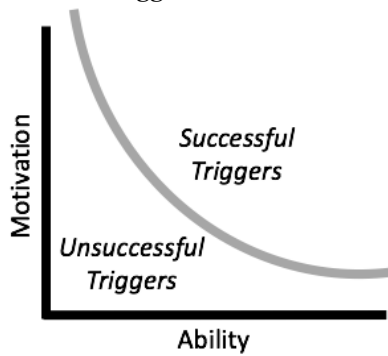

*Figure 2: The Fogg Behavioural Model (adapted from Fogg (2009a))*

Although the FBM and Hook models share a common component (a *trigger*), they serve different levels of the design process. The FBM allows designers to consider the higher level constructs of ability and motivation in system design, whereas the Hook model is better suited to designing specific features. Therefore, we will be using a mix of both models when designing our persuasive technology in this research, in order to carry out a more rigorous design process.

## 2.4 Related Work

To develop the persuasive system for this research, we have elected to follow an 8-step process (Fogg 2009b) that is shown in Figure 2. As part of this process it is important to find relevant examples of persuasive systems (step 5 in Figure 3). This is to ensure that past mistakes can be avoided and more importantly, past successes can be replicated.

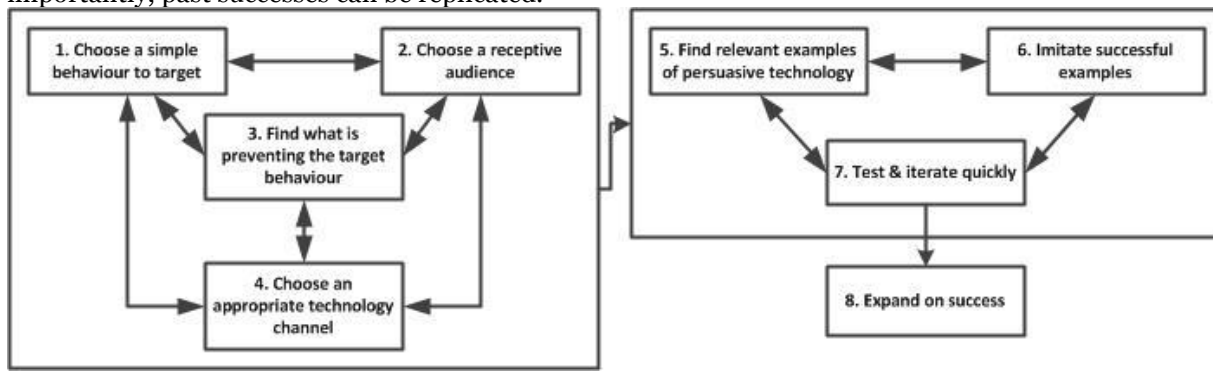

*Figure 3: 8-step Design Process for Persuasive Technology (adapted from Fogg (2009b))*

A similar example of persuasive technology that is related to this research is of a system designed to persuade users to create stronger passwords, called Persuasive Text Passwords (Forget et al. 2008). The problems in password creation and study habits do have some similarity. That is, people are typically aware that they should create strong passwords, however they lack effort and motivation to do so. The investigators were able to influence users to repeatedly form better passwords by using a method called "tunnelling". This involves guiding the user through the process they wish to influence in an incremental way. This may be a useful technique to encourage good study habits, particularly for students who are not aware of what a good study habit might entail.

Another relevant example of persuasive design, not directly in the form of technology, can be found in the Persuasive Trash Cans study (De Kort et al. 2008). The study aimed to uncover the differences between implicit and explicit activation of behavioural norms by placing signs with text to elicit varying types of activation. It was found that explicit activation (rather than implicit) was clearly more likely to result in a change of behaviour, even if the implicit norm for the individual was to avoid littering. This too draws parallels to study habits in the sense that a student's norm may be to study properly, but without explicit activation of that behaviour they are less likely to do so. Therefore, our system should make the study habits explicitly aware to the students. Finally, visual feedback has also been shown to be an effective tool to influence behaviour. The Waterbot system was designed to improve safety, hygiene and water conservation at the sink (Arroyo et al. 2005). It achieves this by providing a variable schedule of reinforcement about the behaviours around the use of the sink it was targeting.

## 3　　System Requirements

This section outlines what is required of the system from a theoretical standpoint. We will progress through the second and third stages of the design process outlined in the Background section. Firstly, the persuasion context is analysed. The requirements of the system qualities will be analysed followed by the selection of the target behaviour and appropriate technology channel.

### 3.1　　The persuasion context

This research is concerned with study habits and so this places the general context for the system in the learning environment. However, what constitutes the learning environment is ever changing. Traditionally, the classroom was the main setting in which most of a student's learning would take place. However, with the rise of computationally powerful mobile technology, learning can take place anytime and anywhere. In fact, bite-sized learning is quickly becoming students' preferred delivery method for learning (van der Meer et al. 2015), by giving students the ability to break down their learning into smaller manageable pieces. Despite this change in the learning environment landscape, the constant is that there are still two prominent actors in education. That is, the student and the instructor. This implies that the focus of analysing the *intent, event* and *strategy* of persuading the student's behaviour should be on the people involved and not necessarily the physical setting and location of where it is taking place.

Understanding who the persuader is and what the change type is forms the *intent* aspect of persuasion. The key decision that needs to be evaluated at this point is whether or not the change type is about attitude or behaviour change. For this research, the emphasis is on behaviour change, as it is the study behaviours that students exhibit that we are aiming to adjust. The persuader is the instructor of the class. Although the instructors will not be physically performing the persuasion themselves, they are doing so vicariously through the system. In regard to providing *system credibility*, the student will need to make the association that the system is instructor-led by attending classes, such as lectures or tutorials and being told about the availability of the system recommending its use. An important consideration that needs to be made about the use of the app concerns the *event* of the persuasion. For persuasion to occur, the system will need to align in some way with the student's goals, such as their desired grade level. Finally, the strategy that will be used to carry out the persuasion is critical. The content and format of the *message* needs to be carefully crafted to improve the chance of persuasion taking place. Positive messages need to be sent to the student encouraging them to continue their effort towards altering their study behaviour.

### 3.2　Target behaviours

The primary purpose of building a persuasive system is to persuade people into altering their behaviour. To do this, a target behaviour (or a set of behaviours) needs to be selected. For this research, we will be selecting target study habits from the three models of academic performance outlined in section 1. The habits were selected on the basis of being feasible to implement in a mobile app. For instance, a habit such as "*During class time I often miss important points because I'm thinking of other things*" would be a difficult habit to target in terms of specific features, and would require an entire system-level approach to be effective. Therefore, the study habits we wish to target are as follows:

(1) I usually study in a place where I can concentrate on my work, (2) I find it hard to stick to a study schedule, (3) when I study for a class I pull together information from different sources, such as lectures, readings and course materials, (4) when I study for a subject, I write brief summaries of the main ideas from the readings and my class notes, (5) when I study for a subject, I often set aside time to discuss material with a group of students from the class, (6) when studying for a subject, I often try to explain the material to a classmate or friend.

Observing the six study habits selected, it is possible to categorise them into more general behaviours. That is, habits 1 and 2 can be considered as *scheduling*, 3 and 4 can be considered *preparation for class*, and 5 and 6 can be considered *group study support*. The literature suggests targeting only a single behaviour, however, these study habits logically pair together. By grouping the habits into general behaviours, we can still target a single type of behaviour at any point in time, while dealing with several specific habits. This also helps to structure our potential approach to longer-term development plans for the system. Scheduling is the prerequisite target behaviour from which any other study habit interventions can build. Without a study schedule, it will be difficult to develop features such as taking notes after class, or arranging study sessions with classmates, as this information will not be known to the system. Therefore, by establishing the scheduling behaviours first, future behavioural targets will be easier to develop, and more likely to succeed in persuading the student's behaviour.

## 4 System Design

Following the establishment of the requirements for a prototype system and outlining a structure to the possible study habits that will be targeted, more specific details can now be designed about the actual design of the system. This section will begin by analysing an existing learning tool platform, and then outline the design for a potential persuasive mobile app. The overall persuasive system will be formed by the combination of both the existing learning tool and a mobile app.

### 4.1 Platform Selection

We are particularly interested in a learning tool called TTM (Task-Test-Monitor), which is currently being used by several courses in a Business Information Systems degree at an Australian university (Grigoriou et al. 2015). TTM is a web-based tool that is used during tutorials where students can download some task instructions, carry them out, and then complete a short multiple-choice test to verify their understanding of the concept to be learnt in the task. TTM addresses an issue across a number of information systems development courses: students are not completing the necessary tutorial work before their assessments. From the students' perspective however, the focus of the system is on providing feedback for their work and monitoring overall progression. Students have unlimited attempts at completing multiple-choice tests, allowing them to explore all possible feedback and can monitor their performance through various charts and graphs. They can also cycle through their historical attempts at each test for revision. TTM was generally designed as a test-bed from which other applications can be built as layers on top of it, hence, TTM is a suitable candidate on which to build a new component, with that being a mobile app.

### 4.2 Technology channel

The fourth step in the eight step persuasive design model focuses on selecting an appropriate technology channel. The TTM system described in the previous section uses the web as its technology channel. Although this is a useful channel for reaching undergraduate students, it is not easily integrated into their daily lives. That is, it requires the user to manually visit the mobile web site or bookmark it for interaction to occur. Alternatively, native mobile apps have the same level of portability, with the added advantage of being able to trigger users through the use of push notifications and updates. Therefore, a mobile app will be the technology channel of choice for this research. TTM will be used as the general platform and the mobile app will serve as a companion to it. This direct coupling of the app with a system that students' use for their tutorials, as well as the deliberate design to be persuasive, is what is expected to provide an advantage over using other standalone study apps.

### 4.3 App features

The prototype persuasive mobile app that is to be built will have features that are designed to support the target behaviours outlined in section 3.2. This section will describe the potential features and how they will follow the Hook model to increase the likelihood of habit adoption.

#### 4.3.1 Target behaviour: Scheduling

*Scheduling* forms the base from which the other features will build upon. The features of the app will need to ensure that they support students to create a schedule and then working with the students to continue to maintain and adjust it as necessary. The two underlying habits we are targeting here are:

1. I find it hard to stick to a study schedule.

2. I usually study in a place I can concentrate on my work.

Organising a study schedule is an issue of motivation but for competing reasons. For some students, planning may not occur as students simply want to complete their work with minimal effort. Planning adds extra work, which they may not want to do. Alternatively, students may enthusiastically plan schedules that remain static, but then are not followed for very long as the motivation to do so declines. Scheduling is not an issue of ability as it is a fairly simply task. The motivation to follow through is the key issue.

We propose three features to improve students scheduling habits for study. The general process for how these features will operate in terms of the Hook model can be found in Table 2.

1. A wizard that helps students to plan out their week and find time to study.

Upon opening the app for the first time, students will be asked to create a weekly schedule for study. They will be required to enter their class timetable, external work commitments and any other regular and relevant events (such as sport or other interests). The app will then ask the students if they typically prefer to work earlier or later in the evening, so as to ease them into starting their habit alteration. The app will then analyse the students' weekly commitments and suggests a time when they may be able to commit to a weekly study session. This will begin with one session per class every week, and over time, based on class performance in TTM, the app may suggest students to add more study sessions for a particular class.

2. Check-in to a study session.

Having committed to a weekly study session, the students will be notified via push notifications or email that their scheduled study time is about to commence. The students will need to "check-in" to the session and at the end, the app will notify the students and ask them to "check-out". To do so, they will need to rate how effective the session was and the environment in which they studied in. This data will be used as feedback to help adapt the personalised schedule to the students' needs.

3. Random suggestions for good places to study.

The app will monitor how well the students have been able to find places to concentrate while studying, based on their rating of the study environment in each session. If students have repeatedly struggled to concentrate, random suggestions will be sent to them such as library reading rooms, or typical quiet locations that are generally conducive to concentration.

| Hook Process ||||
|---|---|---|---|
| **Trigger** | **Action** | **Variable Reward** | **Investment** |
| Student is greeted with wizard setup screen on first run of app. Student is sent random notifications with suggestions on places to study. | Student will enter their schedule data and then rate their study sessions when checking-in. | Visual indicator of how well the student is keeping to their schedule | Entering scheduled data and rating the sessions. |

*Table 2. Hook process for Scheduling habit*

### 4.3.2 Target behaviour: Preparation for class

Reading prescribed text and practising relevant skills prior to attending class helps a student become actively involved in the classroom, particularly if the course uses a flipped classroom approach. The underlying habits targeted for this feature are:
- When I study for a class I pull together information from different sources, such as lectures, readings and course materials.
- When I study for a subject, I write brief summaries of the main ideas from the readings and my class notes.

Sourcing information is primarily an issue of motivation. Many sources are easily available to students on the Internet. Taking notes is also a question of motivation as there is nothing difficult about doing so, many students may simply opt not to do so. Possibly due to the content still being fresh in their minds, they may not feel the pressing need to capture that. The issue of ability to seek sources is only in information literacy. That being, students potentially not knowing which sources to trust or how many are required to prepare adequately. Taking notes is a similar scenario, where students may not know how to write effective notes and hence may not do it.

We propose two features to improve students scheduling habits for study. The general process for how these features will work as per the Hook model can be found in Table 3.

1. A checklist that helps students to ensure they have sourced enough info before commencing a task.
2. Timely notifications to write summary notes right after class while the concepts are still fresh.

Before commencing required tasks, the app presents a series of preparatory tasks to complete, including (if applicable): reading lecture and tutorial notes, relevant textbook passages, any online links shared by the lecturer or tutor as well as the students' personal notes from the previous week. The app will also encourage further reading by suggesting students find their own interesting links or articles. A reminder will be sent to students prior to their class for the week, with ample time to carry out the preparatory reading. At the conclusion of class, students will be prompted by notification to complete summary notes of what was just covered.

| Hook Process | | | |
|---|---|---|---|
| **Trigger** | **Action** | **Variable Reward** | **Investment** |
| Weekly notification of "this week's reading list" on the first screen shown in the app. | Tick off each source the student has used. | Colour change of completion bar and positive message displayed. | Visual progress towards completion. |

*Table 3. Hook process for Preparation habit*

### 4.3.3 Target behaviour: Group study

Group study can be very beneficial to one's learning progress. It provides the opportunity to share new ideas and discuss material to gain clarity of understanding. Few students are proactive in organising robust study groups and many degenerate into a social gathering. These features will overcome this issue by matching suitable students together tracking the effectiveness of the group study session. The specific habits being targeted are:

- When I study for a subject, I often set aside time to discuss material with a group of students from the class.
- When studying for a subject, I often try to explain the material to a classmate or friend.

Socialising is not usually an issue of motivation to establish it, but a motivation to persist with a planned schedule. The rating of the other students' contribution to the group may help alleviate this. Organising a group study session often requires effort to align schedules. The app already has student schedules so it helps to semi-automate this process, reducing ability required.

### 4.3.4 Features

We propose two features to improve students scheduling habits for study. The general process for how these features will work as per the Hook model can be found in Table 4.

1. Invite friends to join study session. The app allows the student to see a list of students from the same class (if the other students provided the information in the schedule feature).

The app will suggest to students who may be beneficial for them to work with. For example, if a student is struggling in web programming, then someone with strong web programming test results will be suggested. If the student elects to invite others to a study session, at the end of the session the student can rate how well he or she studied with the others.

2. Suggestions to explain concepts to friends in the study session.

This feature works in reverse to inviting friends to join a study session. Students will be automatically suggested to work with others in pairs, with the system identifying one student who is a higher performing student and one who is lower performing. The students will be unaware of which they are as each will be asked to attempt to discuss concepts with their partner in the same manner.

| Hook Process | | | |
|---|---|---|---|
| **Trigger** | **Action** | **Variable Reward** | **Investment** |
| After a series of effective study sessions, the app makes a suggestion to invite friends | The student needs to look through the list and select other students known to him or her. | Students who have studied with the student can be endorsed as "helpful" which is sent to other students' apps. | There is a social investment in forming the study group. As time progresses, the student may feel compelled by the duty of working with the group. |

*Table 4. Hook process for Group Study habit*

# 5　Future Work

With the first six steps of the persuasive design process being carried out, the next step is to test and iterate quickly. A prototype of the mobile app presented in this paper will be constructed. The goal is to implement the scheduling features discussed in section 4.3.1 and continue to iterate the development throughout the course of the semester. The scheduling features will be adjusted based on feedback from students and new features (such as the planning and group study features) will be implemented over time. Upon completion of the semester data will be collected and analysed and an evaluation of whether the system was successfully able to alter the study habits of students will be performed.

Once the app is implemented, students will be given the opportunity to download the app in the first week of semester and create their study schedule, with the initial goal being to find at least one session per week of dedicated study. Students can then check into to a session and rate how effective they believe their session was. The app will then adjust the schedule based on this feedback. To improve group study, once the student has found a schedule that works for him or her individually, the app will recommend other students from the class to join him or her. Finally, to influence preparation for class, students will be sent reminders before and after class to ensure they have read the required text, and take summary notes after class.

# 6　Summary

Altering the study habits of undergraduate students can be difficult. Technology can be used to help alleviate these difficulties by providing a means to influence behaviours of individuals. Various disciplines have shown that technology is quite adept at achieving this, particularly e-commerce companies. Persuasive technology works best when there is a clear behaviour that is being targeted. To inform the design of a persuasive system for improving study habits, we selected the habits of study scheduling, class preparation and group study as the focus of our design. To implement this, we selected an existing learning tool in TTM and designed features for a mobile app that will form a new component to the existing TTM system. The combination of TTM and the proposed mobile companion app form the overall persuasive system which will be used influence student study habits to improve their learning outcomes.

# 7　References

Arroyo, E., Bonanni, L., and Selker, T. 2005. "Waterbot: Exploring Feedback and Persuasive Techniques at the Sink," *Proceedings of the SIGCHI conference on Human factors in computing systems*: ACM, pp. 631-639.

Chen, I.J., and Popovich, K. 2003. "Understanding Customer Relationship Management (Crm): People, Process and Technology," *Business process management journal* (9:5), pp 672-688.

Credé, M., and Kuncel, N.R. 2008. "Study Habits, Skills, and Attitudes: The Third Pillar Supporting Collegiate Academic Performance," *Perspectives on Psychological Science* (3:6), pp 425-453.

De Kort, Y.A., McCalley, L.T., and Midden, C.J. 2008. "Persuasive Trash Cans: Activation of Littering Norms by Design," *Environment and Behavior*).

Eyal, N., and Hoover, R. 2013. *Hooked: A Guide to Building Habit-Forming Products*. Createspace Independent Pub.

Filippou, J., Cheong, C., and Cheong, F. 2015. "Designing Persuasive Systems to Influence Learning: Modelling the Impact of Study Habits on Academic Performance," in: *PACIS 2015*. Singapore.

Fogg, B. 2002. "Persuasive Technology: Using Computers to Change What We Think and Do," *Ubiquity* (2002:December), p 5.
Fogg, B. 2009a. "A Behavior Model for Persuasive Design," *Proceedings of the 4th international conference on persuasive technology*: ACM, p. 40.
Fogg, B. 2009b. "Creating Persuasive Technologies: An Eight-Step Design Process," *Proceedings of the 4th International Conference on Persuasive Technology*, p. 44.
Forget, A., Chiasson, S., van Oorschot, P.C., and Biddle, R. 2008. "Persuasion for Stronger Passwords: Motivation and Pilot Study," in: *Persuasive Technology*. Springer, pp. 140-150.
Grigoriou, T., Cheong, C., and Cheong, F. 2015. "Improving Quality of Feedback Using a Technology-Supported Learning System," in: *PACIS 2015*. Singapore.
Hamari, J., Koivisto, J., and Pakkanen, T. 2014. "Do Persuasive Technologies Persuade?-a Review of Empirical Studies," in: *Persuasive Technology*. Springer, pp. 118-136.
Kaptein, M., Markopoulos, P., de Ruyter, B., and Aarts, E. 2009. "Can You Be Persuaded? Individual Differences in Susceptibility to Persuasion," in: *Human-Computer Interaction–Interact 2009*. Springer, pp. 115-118.
Lockton, D., Harrison, D., and Stanton, N.A. 2010. "The Design with Intent Method: A Design Tool for Influencing User Behaviour," *Applied ergonomics* (41:3), pp 382-392.
Oinas-Kukkonen, H., and Harjumaa, M. 2009. "Persuasive Systems Design: Key Issues, Process Model, and System Features," *Communications of the Association for Information Systems* (24:1), p 28.
Pintrich, P.R. 1991. "A Manual for the Use of the Motivated Strategies for Learning Questionnaire (Mslq),").
Pintrich, P.R., Smith, D.A., García, T., and McKeachie, W.J. 1993. "Reliability and Predictive Validity of the Motivated Strategies for Learning Questionnaire (Mslq)," *Educational and psychological measurement* (53:3), pp 801-813.
Saari, T., Ravaja, N., Laarni, J., Turpeinen, M., and Kallinen, K. 2004. "Psychologically Targeted Persuasive Advertising and Product Information in E-Commerce," in: *Proceedings of the 6th international conference on Electronic commerce*. Delft, The Netherlands: ACM, pp. 245-254.
van der Meer, J., Berg, D., Smith, J., Gunn, A., and Anakin, M. 2015. "Shorter Is Better: Findings of a Bite-Size Mobile Learning'pilot Project," *Creative Education* (6:03), p 273.

## Copyright